\documentclass[a4paper,11pt]{article}
\usepackage{ILD}
\usepackage{subcaption}

\usepackage{amssymb}
\usepackage{amsfonts}
\usepackage{amsmath}

\usepackage{times}
\usepackage[T1]{fontenc}

\usepackage{lineno}

\newcommand\onbehalf{\vskip 1.5in 
			 \vbox{\small
			   \hskip -0.23in 41st International Conference on High Energy physics - ICHEP2022\\
 6-13 July, 2022\\  Bologna, Italy \\}}

\title{Stau searches and measurements with the ILD concept at the International Linear Collider}

\ildproc{phys}{2022}{010}
\date{\today}
\addauthor{M.T. N{\'u}{\~n}ez Pardo de Vera\footnote{On behalf of the ILD Concept Group}\footnote{Speaker}}
{\institute{1}}
\addauthor{M. Berggren}
{\institute{1}}
\addauthor{J. List}
{\institute{1}}

\addinstitute{1}{  Deutsches Elektronen-Synchrotron DESY,\\
                      Notkestr. 85, 22607 Hamburg, Germany \\
                    \vskip 0.1in
                    E-mail: maria-teresa.nunez-pardo-de-vera@desy.de}

\abstract{One of the most interesting channels to search for SUSY is the direct pair-production of
  the $\tau$-lepton superpartner, $\widetilde{\tau}$. The $\widetilde{\tau}$ is with
  high probability the lightest of the scalar leptons, so one of the first SUSY particles that can
  be observerd, and the signature of $\widetilde{\tau}$ pair production signal events is one of the
  most difficult ones, yielding to the ``worst'' and so most global scenario for the searches.
  Analysis performed at LEP set the current model-independent $\widetilde{\tau}$ limits, suffering
  from the low energy of this facility. Only under strong model assumptions, these limits are extended to
  higher masses by LHC studies.
  In this contribution we show the capability of the ILC, a future electron-positron collider with
  energy up to 1 TeV, for determining $\widetilde{\tau}$ exclusion/discovery limits in a
  model-independent way, including an overview of the current state-of-the-art.
  The determination of the ``worst'' scenario for $\widetilde{\tau}$ exclusion/discovery, taking
  into account the effect of the $\widetilde{\tau}$ mixing on $\widetilde{\tau}$ production
  cross-section and efficiency, is also presented.
  For selected benchmarks, the prospect for measuring masses
  and polarised cross-sections will be shown. The studies were done studying events passed through the
  full detector simulation and reconstruction procedures of the International Large Detector (ILD) concept
  at the ILC. The simulation included all SM backgrounds, as well as the machine induced ones.

  \onbehalf

  }

\begin{document}
\titlepage

\section{Introduccion and limits at other facilities}
\begin{figure}[htbp]
    \centering
    \subcaptionbox{}{\includegraphics [scale=0.30]{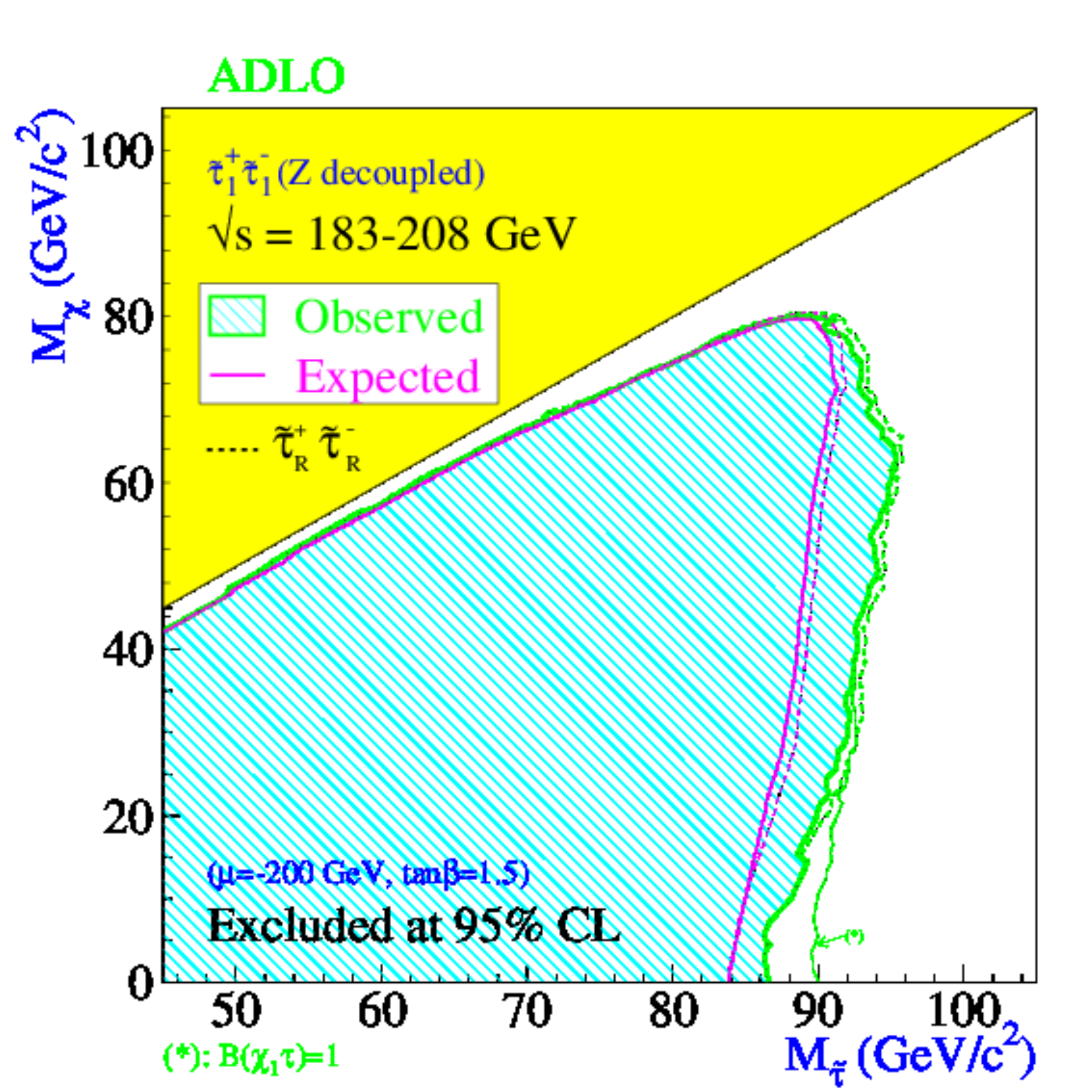}}
    \subcaptionbox{}{\includegraphics [bb=252 178 807 563,clip=true,scale=0.35]{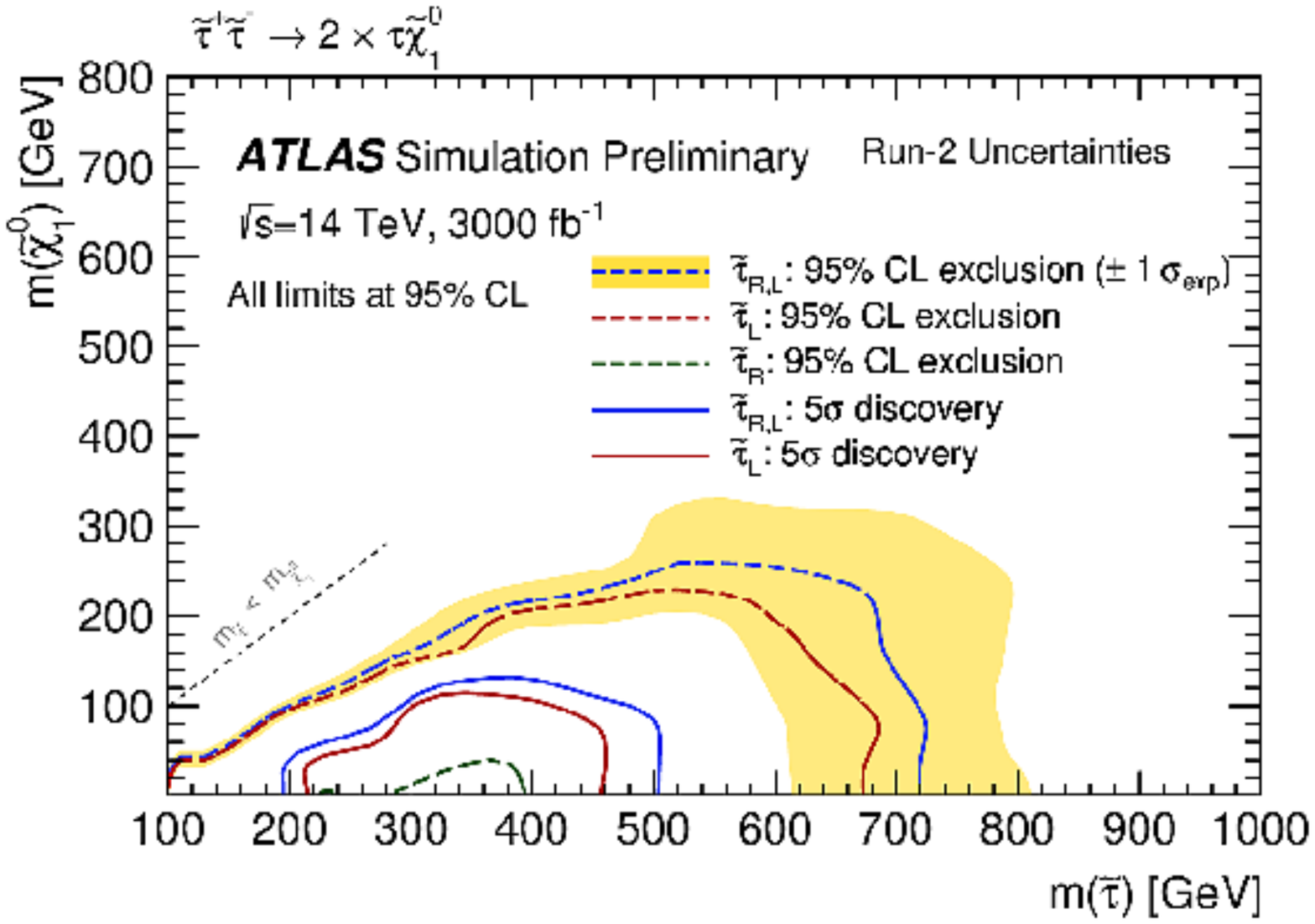}}
    \caption{(a): 95$\%$ CL exclusion limits for $\widetilde{\tau}$ pair production obtained combining data collected at the four LEP experiments with
      energies ranging from 183 GeV to 208 GeV. From~\cite{LEPSUSYWG/04-01.1}.
       (b): 95$\%$ CL exclusion and discovery potential for $\widetilde{\tau}$ pair production at the HL-LHC, assuming
    $\widetilde{\tau}_{L}^{+}\widetilde{\tau}_{L}^{-}$ + $\widetilde{\tau}_{R}^{+}\widetilde{\tau}_{R}^{-}$ production, $\widetilde{\tau}_{L}^{+}\widetilde{\tau}_{L}^{-}$ production or $\widetilde{\tau}_{R}^{+}\widetilde{\tau}_{R}^{-}$ production. From~\cite{ATLAS:2018diz}.}
    \label{lep_lhc_limits}
  \end{figure}

  Two important conditions when searching for SUSY at future facilities are look for
  the lightest accesible particle in the SUSY spectrum and cover the most difficult
  scenario. Both of them are satisfied by the $\widetilde{\tau}$.
  As a consequence of the mixing of both $\widetilde{\tau}$
  weak hyper-charge states, $\widetilde{\tau}_{L}$ and $\widetilde{\tau}_{R}$, it is expected that the lightest of
  its physical states, $\widetilde{\tau}_{1}$, will be the lightest slepton.
  This mixing also points out to a lower cross-section: the strength of the
  $Z^{0}$/$\gamma$ $\widetilde{\tau}$ $\widetilde{\tau}$ coupling depends on the $\widetilde{\tau}$ mixing,
  reaching its minimum value when the coupling $\widetilde{\tau}_{1}$ $\widetilde{\tau}_{1}$ $Z^{0}$ vanishes.
  A difficult experimental signature is due to the fact that its SM partner is unstable, decaying before it
  can be detected, and, as a further complication, some of its decay products are undetectable neutrinos.
  One can conclude that any other NLSP would be easier to find than the $\widetilde{\tau}$ and, therefore,
  $\widetilde{\tau}$ production studies might be seen as the way to determine the guaranteed discovery or
  exclusion reach for SUSY.
  $\widetilde{\tau}$ studies are also theoretically motivated: the observed relic density could be
  accomodated with a light $\widetilde{\tau}$ due to an enhanced $\widetilde{\tau}$-neutralino coannihilation.
  Figure~\ref{lep_lhc_limits}(a) shows the $\widetilde{\tau}$ mass limits from LEP experiments~\cite{LEPSUSYWG/04-01.1},
  being the most model-independent ones. Depending on the mass difference between the $\widetilde{\tau}$ and
  the neutralino, the minimum value of the $\widetilde{\tau}$ mass ranges from 87 to 93 GeV. The limits are valid
  for any mixing and any value of the model-parameters not shown in the plot. 
  
  A $\widetilde{\tau}$ mass below  26.3 GeV, for any mixing and any mass difference larger than the $\tau$ mass, is
  excluded by an analysis from the DELPHI experiment targeted at low mass differences~\cite{Abdallah:2003xe}. 

  The limits on the $\widetilde{\tau}$ mass determined at the LHC are only valid under certain assumptions:
  ATLAS and CMS assume $\widetilde{\tau}_{R}$ and $\widetilde{\tau}_{L}$ to be mass-degenerate and without mixing.
  The future HL-LHC should provide an improvement on the $\widetilde{\tau}$ limits. Indeed simulation studies have
  already been performed in both experiments~\cite{ATLAS:2018diz,CMS:2018imu}, providing upper limits for $\widetilde{\tau}$ masses
  increased by about 300~\,GeV with respect to the ones from the previous studies, but still suffering from the same constraints.
  ATLAS limits for pure $\widetilde{\tau}_{R}$ pair production, that could be considered
  the closest case to the physical lightest $\widetilde{\tau}$ since it is likely to be the lightest of the two
  weak hyper-charge states and the one with the lower cross section,
  show no discovery potencial (see figure~\ref{lep_lhc_limits}(b)), only exclusion one.
  Neither discovery nor exclusion potential is shown for the scenario allowing
  $\widetilde{\tau}$-neutralino co-annihilation. 
  
  \section{Signal and background}
  This study assumes R-parity conservation, the $\widetilde{\tau}$ as the NLSP, and mass differences
  above the mass of the ${\tau}$. Under these conditions, $\widetilde{\tau}$'s
  will be produced in pairs via $Z^{0}$/$\gamma$ exchange in the s-channel and they will decay to
  a ${\tau}$ and an LSP. The ${\tau}$ will decay before leaving any signal in the detectors while
  the LSP will leave the detector without being detected.
  The only detectable activity in the signal events is therefore
  the visible decay products of the two ${\tau}$'s. 
  These signal events are then characterised by a large missing energy and momentum (due to invisible LSPs and
  neutrinos from both ${\tau}$-decays), large
  fraction of the detected activity in the central region of the detector ($\widetilde{\tau}$'s are scalars),
  un-balanced transverse momentum, large angles between the two $\tau$-lepton directions and zero forward-backward asymmetry
  (direction of the $\widetilde{\tau}$ does not  strongly correlate to that of the visible $\tau$ after the  decay).
  The main sources of background  are SM processes with real or fake missing energy.
  They can be classified into ``irreducible'' and ``almost irreducible'' sources. The first are events with two $\tau$'s
  and neutrinos, i.e. real missing energy, being the main contribution $ZZ$ events
  with one $Z$ decaying to two neutrinos and the other to two $\tau$'s, and leptonic $WW$ events,
  where both the $W$'s decays to $\tau$ and neutrino.
  The second group of events are those looking after reconstruction very similar to two $\tau$'s and neutrinos,
  mainly events with two soft $\tau$-jets, with two other leptons plus true
  missing energy or with two $\tau$'s plus fake missing energy.

  \section{Analysis and limits}
\begin{figure}[htbp]
    \centering
    \subcaptionbox{}{\includegraphics [width=0.45\textwidth]{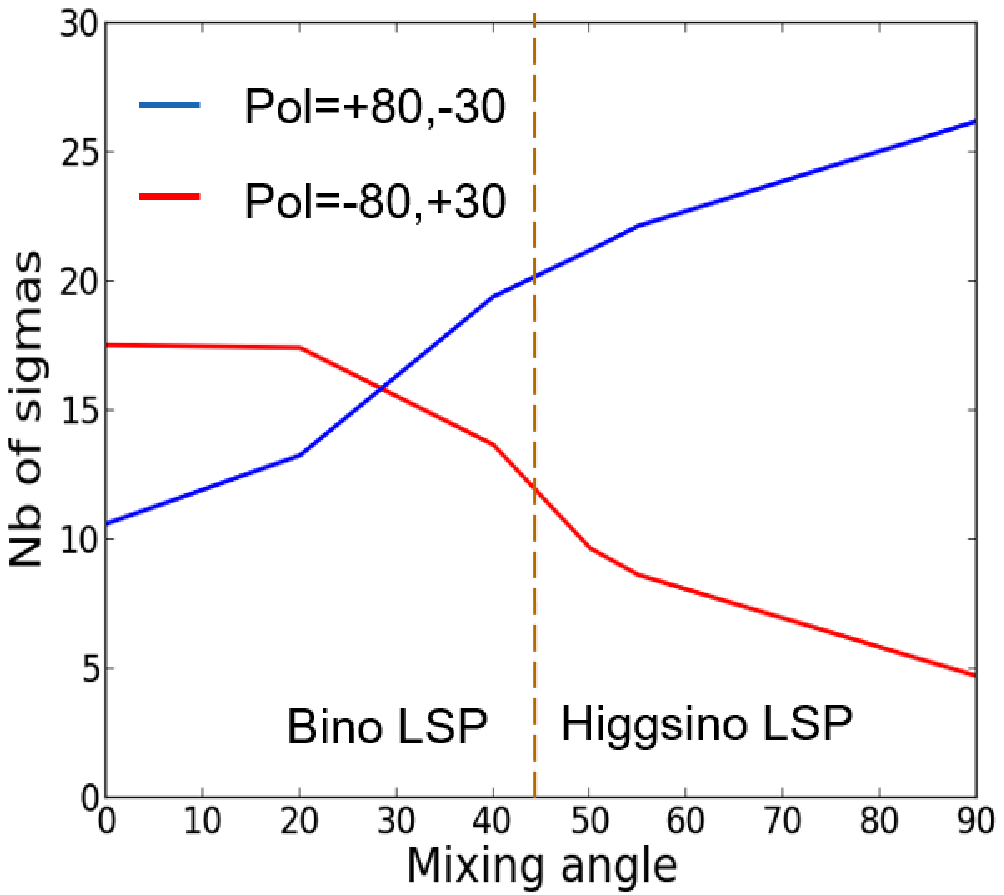}}
     \subcaptionbox{}{ \includegraphics [width=0.45\textwidth]{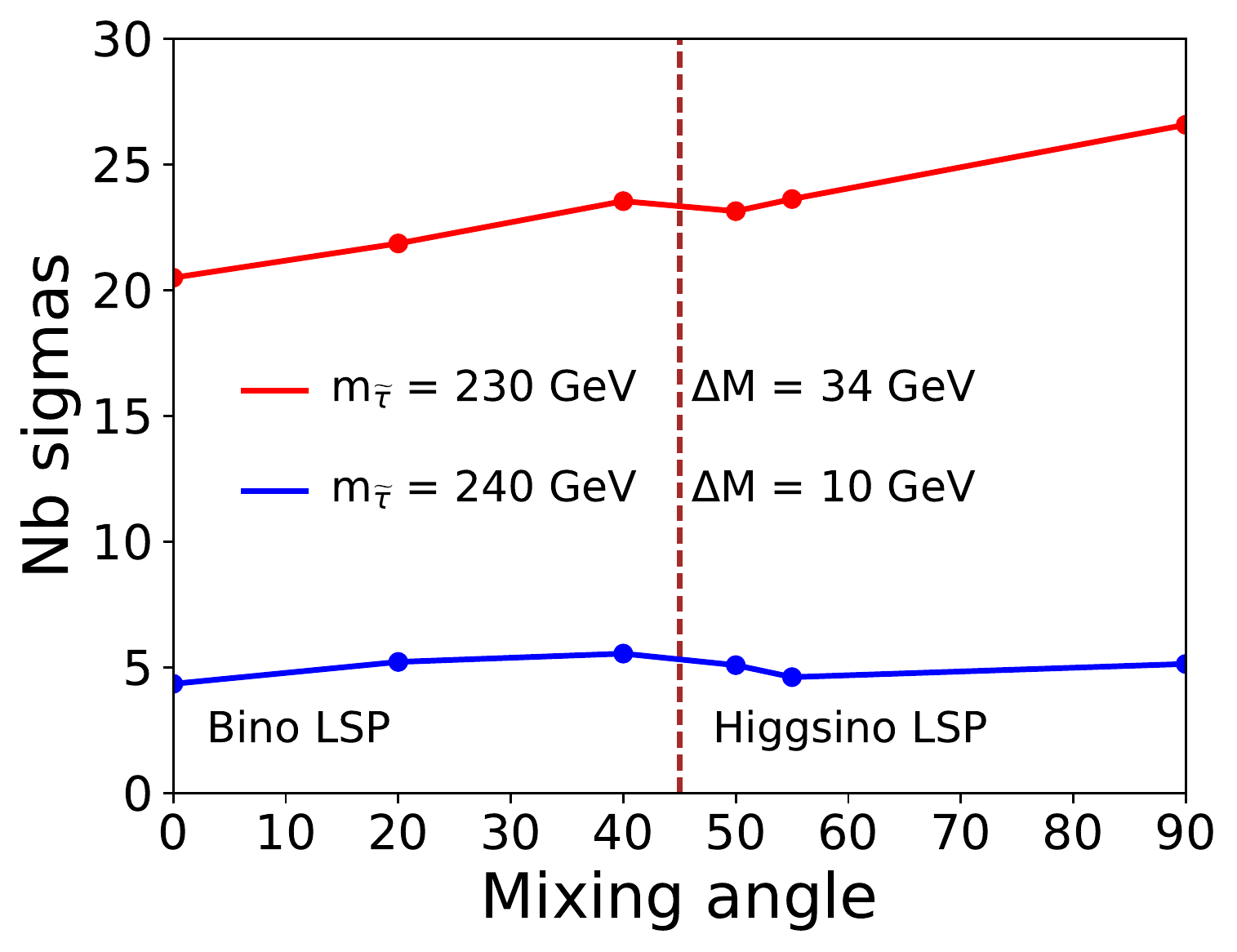}}
   \caption{(a): Signal over background significance as a function of the $\widetilde{\tau}$ mixing angle for both main ILC polarisations.
    (b)Signal over background significance weighting both polarisations using the likelihood-ratio static in the H-20 ILC conditions.}
    \label{sigmas_mixings_weighted}
  \end{figure}

  According to the H-20 running scenario for the ILC500~\cite{staupaper}, an integrated luminosity of 1.6~\,ab$^{-1}$
  at $\sqrt{s}=500$\,GeV for each of the beam polarisations, $P(e^{-},e^{+})=(+80\%,-30\%)$ and $P(e^{-},e^{+})=(-80\%,+30\%)$,
  was assumed in the study.  
  The ``worst'' scenario for $\widetilde{\tau}$ exclusion/discovery was analysed taking into account the effect of
  the $\widetilde{\tau}$ mixing on $\widetilde{\tau}$ production cross-section and efficiency.
  Figure~\ref{sigmas_mixings_weighted}(a) shows the signal over background
  significance as a function of the mixing angle for each polarisation. The final significance
  was computed weighting the contribution of both polarisations by the likelihood ratio statistic.
  Figure~\ref{sigmas_mixings_weighted}(b) plotts the results, showing a rather uniform sensitivity to all mixing angles.
  For the smallest mass differences, the critical ones, a mixing angle around 53$^\circ$, corresponding
  to the lowest cross-section for unpolarised beams, can be taken as the
  worst case for the ILC conditions. 
 
\begin{figure}[htbp]
    \centering
    \subcaptionbox{$\Delta(M) = $ 3 GeV}{\includegraphics [width=0.45\textwidth]{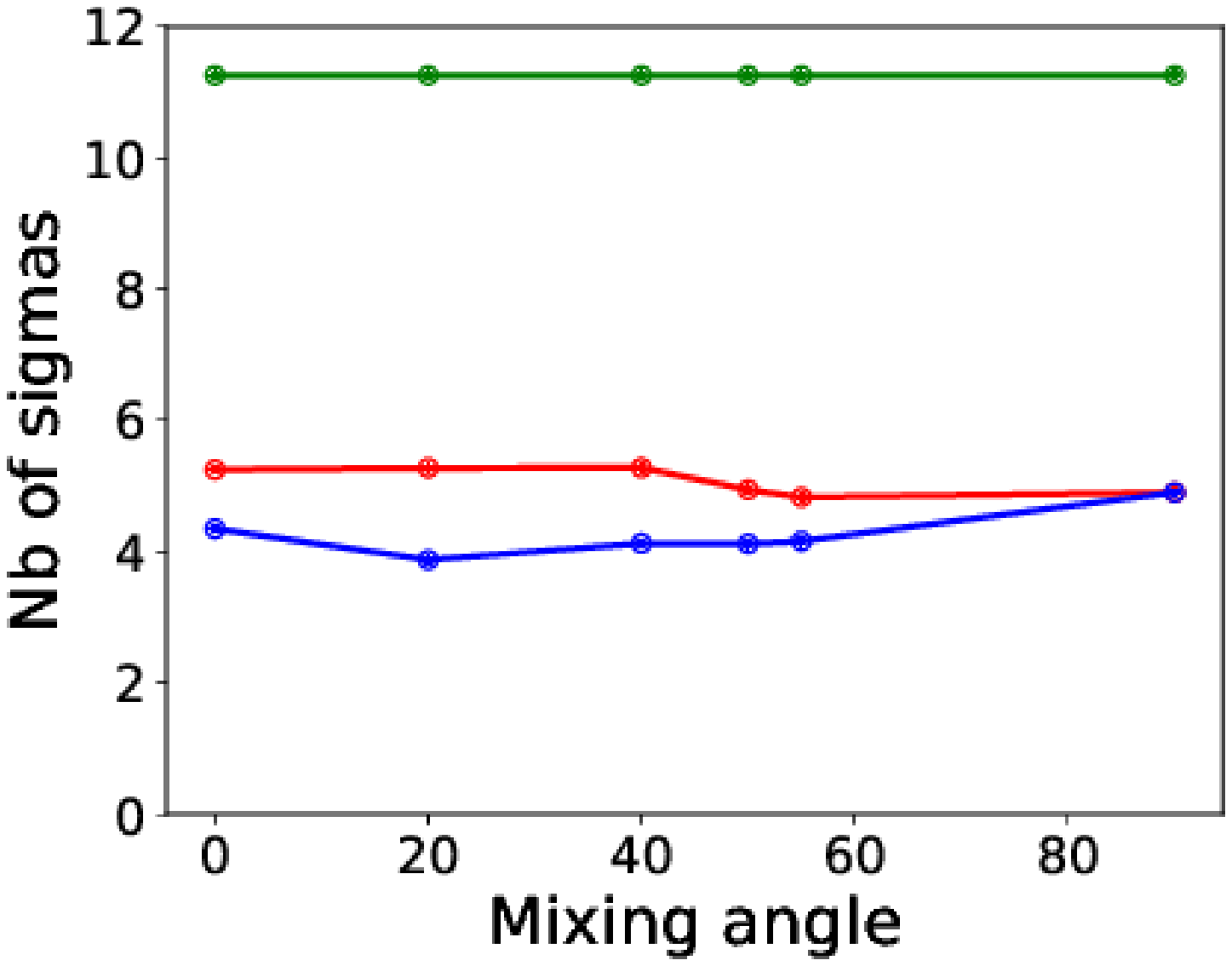}}
    \subcaptionbox{$\Delta(M) = $ 10 GeV}{\includegraphics [width=0.45\textwidth]{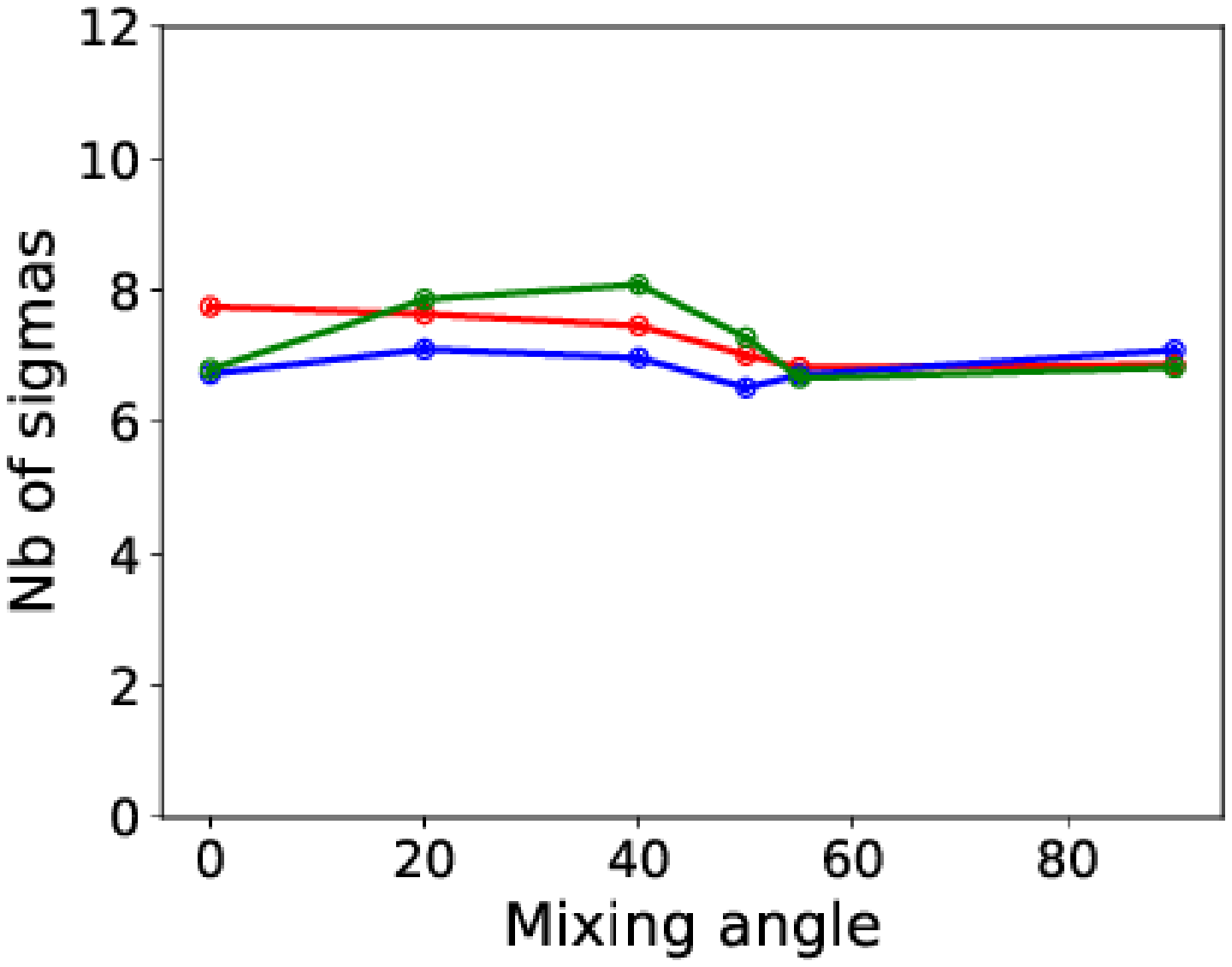}}
    \vspace*{0.4cm}
    \caption{Number of sigmas for a $\widetilde{\tau}$ with mass 240 GeV and different mass-differences.
      The plot assumes H-20 ILC scenario combining both polarisations using the likelihood-ratio
      statistic. Blue lines correspond to the case with all the tracks and the red ones after rejecting tracks
      not satisfying the cuts described in the text. The green curves correspond to the study without overlay
      tracks. 
      }
    \label{nb_sigmas_fullsimu_sgv}
  \end{figure}

  Cuts for separating signal from background have been designed taking into
  account the signal signature and the main background sources.
  The first group of cuts are those in properties that the
  $\widetilde{\tau}$-events {\it must} have, meaning cuts in
  missing energy, visible mass, maximum total momentum and
  maximum momentum of the jets. An algorithm for
  $\tau$-identification was also applied.
  A second group of cuts is based on those properties that the
  $\widetilde{\tau}$-events {\it might} have, but the background
  events will {\it rarely} have, allowing to set cuts requiring
  high missing transverse momentum ($P_{T miss}$), large acoplanarity $\theta_{acop}$,
  large angle to the beam, and large value of the variable $\rho$,
  calculated by first projecting the event on the x-y
  (transverse) plane, and calculating the thrust axis in that plane. The variable $\rho$
  is then the transverse momentum (in the plane) with respect to the
  thrust axis.
  The third group of cuts uses properties of some of the ``almost irreducible'' sources of
  background, using the highly forward-backward asymmetry of $WW$ events with each of the $W$'s decaying to a lepton
  (other than $\tau$) and the visible mass close to $M_Z$ for the $ZZ$ events with one $Z$ decaying to two neutrinos
  and the second one to a electron or muon pair.
  A last cut is based on a property that the signal often {\it does not} have: sizeable energy
  detected at low angles to the beam. 
  After applying these cuts the main sources of remaining background are $WW$ events with each
  $W$ decaying to $\tau\nu$ and events with four fermions in the final state coming from
  $\gamma\gamma$ interactions, mostly $\tau\tau$ events.
  For events with mass difference between 3~\,GeV and the mass of the $\tau$ an additional cut requiring
  an ISR photon was applied.
  Visible tracks from the $\widetilde{\tau}$-decays for small mass differences between $\widetilde{\tau}$ and the LSP
  have very similar properties to the ones of the overlay tracks. For that reason, these tracks can
  not be neglected in this study.
  Overlay tracks are mainly hadrons with low transverse momentum coming from interactions of real or virtual
  photons produced by the beams. An algorithm trying to reduce these overlay tracks was developed based
  on transversal momentum, angular distribution and input parameter significance. 
  Figure~\ref{nb_sigmas_fullsimu_sgv} shows the significance obtained with and without cuts together with the results
  from the {\tt SGV} fast simulations~\cite{staupaper} (without overlay tracks). For the case with the smallest mass difference,
  shown in figure~\ref{nb_sigmas_fullsimu_sgv}(a), 
  there is a strong reduction of the significance when adding overlay tracks.
  For the larger mass-difference ((figure~\ref{nb_sigmas_fullsimu_sgv}(b) ) the degradation is slight.
  In both cases the overlay removal ameliorates the sensitivity.

\begin{figure}[htbp]
    \centering
    \includegraphics [width=0.5\textwidth]{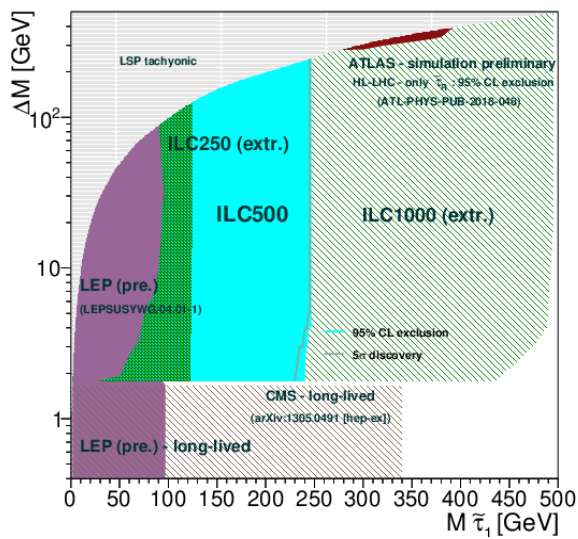}
    \caption{$\widetilde{\tau}$ limits in the $M_{\widetilde{\tau}}$-$\Delta$M plane. ILC results from the current studies are shown together with limits from LEP and LHC. The region with mass differences below the mass of the ${\tau}$ is also shown with LEP and LHC results, even if it is not covered by this study. In addition, the extrapolation of
      the ILC current results to the ILC 250~\,GeV and 1~\,TeV running scenarios is show.}
    \label{exclusion_mstauvsdm}
  \end{figure}

  Figure~\ref{exclusion_mstauvsdm}~\cite{staupaper} shows the projection of the limits in the $M_{\widetilde{\tau}}$-$\Delta$M plane,
  together with the limits from LEP and the projected HL-LHC ones (to be taken with care due to the high model dependence).
  Extrapolation of the ILC limits for the scenarios
  with centre-of-mass energy 250~\,GeV and 1~\,TeV is also shown.
  One can observe that for the ILC exclusion and discovery limits are very close to each other and to the
  ILC kinematic limit.
  The region for mass differences below the mass of the $\tau$ is shown for
  completeness, even if it was not included in this study.

  For specific benchmarks the $M_{\widetilde{\tau}}$ was computed based on the end-point
  of the spectrum from $\tau$ decays and on the $\widetilde{\tau}$ cross-sections,
  achieving per mil-level precision on the measurements.
  ${\tau}$ polarisation and $\widetilde{\tau}$ mixing angle were also computed based on the spectrum
  of the ${\tau}$ decays and $\widetilde{\tau}$ cross-sections and masses, respectively.
  Percent level precission was reached in those cases~\cite{precmeas1}. 

\section{Outlook and conclusions\label{sec:concl}}

The ILC is presented as a promising scenario for SUSY studies.
At this future facility $\widetilde{\tau}$-pair production could be excluded/discovered
up to a few GeV below the kinematic limit, even in the worst scenario and without model
dependencies.

Well motivated regions of the SUSY parameter space, that would most probably not be
covered by the HL-LHC, could be studied.

The effect of the overlay particles, that can no be neglected, is however mitigated applying
optimized tracks.

If the $\widetilde{\tau}$ exists in the kinematic range of the ILC, precision measurements of $\widetilde{\tau}$
properties could be measured at a few percent level.

\end{document}